\documentclass[english,superscriptaddress,prl,twocolumn]{revtex4}
\usepackage[toc,page,title,titletoc,header]{appendix}
\usepackage[T1]{fontenc}
\usepackage{enumerate}
\usepackage{graphicx}
\usepackage[latin1]{inputenc}
\usepackage{epstopdf}
\usepackage{amsthm}
\usepackage{amsmath}
\usepackage{shapepar}
\usepackage{color}

\usepackage{amssymb, amsbsy, amsthm}
\makeatletter

\usepackage{babel}
\makeatother
\begin{document}
\title{ No Disturbance Without Uncertainty as a Physical Principle}
\author{Liang-Liang Sun$^{*}$, Xiang Zhou\footnote[1]{Co-first author}, Sixia Yu\footnote[2]{email: yusixia@ustc.edu.cn}}
\affiliation{Hefei National Laboratory for Physical Sciences at Microscale and Department
of Modern Physics, University of Science and Technology of China, Hefei,
Anhui 230026}
\date{\today{}}
\begin{abstract}
Finding physical principles lying behind quantum mechanics is essential to understand various quantum features, e.g., the quantum correlations, in a theory-independent manner. Here we propose such a principle, namely, \emph{no disturbance without uncertainty}, stating that the disturbance caused by a measurement to a subsequent incompatible measurement is no larger than the uncertainty of the first measurement, equipped with suitable theory-independent measures for disturbance and uncertainty. When applied to local systems in a multipartite scenario, our principle  imposes such a strong constraint on non-signaling correlations that quantum correlations can be recovered in many cases: i.) it accounts for   the Tsirelson's bound; ii.) it provides the so far tightest boundary for  a  family of the noisy super-nonlocal box with 3 parameters,  and iii.) it rules out an almost quantum correlation from quantum correlations by which   all the previous principles fail, as well as the celebrated quantum criterion due to Navascues, Pironio, and Ac\'{\i}n. Our results pave the way to understand nonlocality exhibited in  quantum correlations from local principles. \end{abstract}


\maketitle

{\it Introduction--- }  Quantum  non-locality   is the most pronounced nonclassical feature of quantum mechanics (QM) \cite{bel}.  Understanding  quantum non-locality   is a fundamental  and challenging task that aims, on the one hand, at searching for  physical principles lying behind  its formalism \cite{sp,ic,wvd,cl, by,ca}, on the other hand, at searching for correlation criteria, i.e., the conditions under which a set of observed correlations admit a quantum mechanical description, that witness the boundaries of quantum correlations emerging from the mathematical structure of QM~\cite{npa,tir}.

The boundary between quantum  and local realistic correlations, i.e., to answer the question of being local or nonlocal, is relatively well characterized by the violations of Bell inequalities.  Yet it is notorious difficult to characterize  the boundary between   quantum and  post-quantum correlations, i.e., to answer the question of why certain amount of non-locality is exhibited in quantum correlations.
Even in the simplest Bell scenario the sufficient correlation criterion is inaccessible. Tsirelson, Landau, and Masanes (TLM) derived the first correlation criterion, which becomes sufficient in the case of correlations with unbiased marginals. Later on,   Navascues, Pironio, and
Ac\'{\i}n (NPA) generalized the TLM condition to the general correlations \cite{tir,15',16',17'} and proposed a systematic method using a heirachy of semidefinite positive programs (SDP) to push the boundaries of quantum correlations. NPA criterion is reproduced when the SDP method stops at the first level \cite{sdp1,sdp}.

In comparison, the  theory-independent line of research,  initiated by Popescu and Rohrlich \cite{pr},  focuses on  finding the physical principles that can account for quantum correlations without resorting to quantum  formalism. Generally, the theory-independent approach is less efficient   than the correlation criterion approach in witnessing the quantum correlations, but they provide us deep understandings on the physical principles lying behind QM \cite{tn1}. Van Dam found the first essential discrepancy between QM and  Popescu and Rohrlich (PR) box, an extremal nonlocal correlation, and noted that the PR box requires only a single bit of classical communication for  all distributed computation tasks~\cite{wvd}.  Since then various physical principles such as Information Causality (IC)\cite{ic}, Global Exclusive  \cite{cl,by},  Macroscopic Locality~\cite{ml}, and Local Orthogonality~\cite{lo}  have been proposed.  These principles are able to  recover the Tsirelson's bound but not the boundaries of  quantum  correlations. For example, none of them can  single out quantum correlations   from  almost quantum correlations (AQC),  which is strictly larger than the set of quantum correlations~\cite{acq}.

All those  principles mentioned above are based on multipartite scenario and constraints  are directly imposed on correlations. Recently, a local principle, namely, Uncertainty-Complementarity Balance Relation (UCBR) \cite{me},  has been proposed as an underlying physical principle for quantum correlations. Instead of Bell scenario, sequential measurement scenario is considered in which  incompatible measurements  are performed in certain order on local systems.
 The balance relation establishes quantitative connections between uncertainty, complementarity and nonlocality to account for the Tsirelson's bound \cite{me}. Even though some addition conditions are assumed in order to derive the Tsirelson bound,  the UCBR principle has shown us a way to understand quantum correlations from some local properties of subsystem.

In this Letter,  we pursue this line of research further by introducing a different local principle, namely, No Disturbance without Uncertainty (NDWU), which claims that for two incompatible measurements performed sequentially on a local system, the disturbance caused by the first measurement to the second measurement is no larger than the uncertainty of the first measurement, with properly chosen theory-independent measures for disturbance and uncertainty. Together with non-signaling conditions, our NDWU principle enables us to derive correlation criterion  that can characterize quantum correlations in general probabilistic theories (GPT).  In the simplest Bell's scenario our NDWU principle outperforms all the criteria found so far as it can account for the Tsirelson's bound, recover the  quantum boundaries for a family of the noisy non-signaling boxes, and single out  AQC  from quantum correlations, in which case our local principle  outperforms both the correlation criteria such as quantum NPA, TLM criteria and those based on correlation principles.

\emph{No disturbance without uncertainty --- }
 We  shall work in the framework of a general probabilistic theory in which the concepts of physical state and sharp measurement (observable) are well-defined.  Generally, a system can be prepared in different states which can be any mathematical structure that enable us to predict the statistics of all possible measurements performed on the system or can be operationally taken  as a black box  with  measurement setting as input and  outcome as output.  A sharp measurement is accurate and repeatable.  More precisely,  a sharp measurement $A$ on a physical state $\mathcal{S}$ would return a definite value denoted by $a$, dependent on which  the system is brought into a unique state~$\mathcal{A}^a$, which is independent of the original state, with a probability $p_{\mathcal{S}}({a|A})$. It is repeatable since the same outcome $a$ will be obtained if we perform the same measurement subsequently.
We assume that there are more than one sharp measurement.

 Uncertainty and disturbance in sequentially performing two incompatible measurements are two characteristics features in a general nonlocal theory \cite{gs} and can be quantified in a theory-independent way. Recently, a balance relation between uncertainty and complementary  \cite{me} has been established in a GPT. Inspired by the UCBR, here we also consider a scenario of sequential measurements $A_{0}\rightarrow A_{1}$ in which a sharp measurement $A_0$ is followed by another sharp measurement $A_1$. Since two sharp measurements are involved, it is necessary to consider the  transfer probabilities
  $$\gamma_{A_{1}^{a'};\mathcal{A}_{0}^{a}}:=p_{\mathcal{A}_{0}^{a}}(a^\prime|A_1)$$  of obtaining outcome $a'$ by measurement $A_{1}$ after the first sharp measurement $A_0$ has been performed on the initial state yielding an outcome $a$ and leaving the system at state $\mathcal A_0^a$. We note that the transfer probabilities do not depend on the initial state since the state $\mathcal A_0^a$ does not.
Before presenting our principle we need theory-independent measures to quantify the uncertainty and  disturbance:
\begin{enumerate}[(1)]
\item The uncertainty of a measurement $A_0$, giving rise to probability distribution $\{p({a|A_{0}})\}$, is quantified by
$$\Delta_{A_{0}}:=\sqrt{1-\textstyle\sum_{a}p({a|A_{0}})^{2}}.$$

\item  The disturbed uncertainty of a measurement $A_0$, which is performed after measurement $A_1$, is quantified by a sum of  uncertainties of measurement $A_0$
$$\textstyle\Delta _{A_0|A_{1}}= \sum_{a'}\Delta_{A_0;\mathcal A_1^{a'}}:=\sum_{a'}\sqrt {1-\sum_{a}\gamma^{2}_{A_{0}^{a};\mathcal{A}_{1}^{a'}}}$$
over all possible states after measurement $A_1$.

\item The disturbance  caused by  measurement $A_{0}$ on the subsequent measurement $A_{1}$ is quantified by
$$D_{A_{0}\rightarrow A_{1}}:=\sum _{a^\prime}|p({a^\prime|A_{1}})-p({a^\prime|A_{0}\to A_{1}})|,$$
where $\{p({a|A_{1}})\}$ is the statistics obtained by measuring $A_{1}$ on the original state and
\begin{equation*}
p({a^\prime|A_0\to A_1})=\sum_a p({a|A_0}) \ \gamma_{A_{1}^{a'};\mathcal{A}_{0}^{a}}
\end{equation*}
 is the disturbed statistics obtained by measuring $A_{1}$ after measurement $A_{0}$.
\end{enumerate}

{\bf Theorem 1 } \emph{  On a quantum system with finite levels, if two sharp measurements are performed in order $A_{0}\to A_{1}$  then it holds the following uncertainty disturbance relation}
 \begin{eqnarray}\label{g2}
\Delta_{A_{0}} \Delta_{A_0| A_{1}}\geq D_{A_{0}\rightarrow A_{1}}.
\end{eqnarray}

The  proof is  presented in the supplemental materials (SM). We note that this uncertainty disturbance relation is formulated in a theory-independent manner as all quantities involved, namely, uncertainty and disturbance, are quantified without referring to quantum theory. This means that it is possible for us to impose relation Eq.(\ref{g2}) in a  GPT as physical principle satisfied by all pairs of sharp measurements and the resulting theory shall include quantum theory as a subset. As a qualitative reading, a nonzero disturbance (right hand side) caused by measurement $A_0$ to the subsequent measurement $A_1$ requires a nonzero uncertainty in measuring $A_0$ both in the original state and in the disturbed states resulting from measuring $A_1$ first. Thus we shall refer this qualitative readings from relation Eq.(\ref{g2}) in a general probabilistic theory to as the principle of  {\it no disturbance without uncertainty (NDWU)} and the quantitative uncertainty disturbance relation Eq.(\ref{g2}) as NDWU relation.

For an example let us consider the simplest physical system with sharp measurements having only two outcomes. We denote by $\langle A_0\rangle=p_{0|A_{0}}-p_{1|A_{0}}$ and $\langle A_1\rangle=p_{0|A_{1}}-p_{1|A_{1}}$ the expectation values of the  two-outcome measurements $A_{0}$  and $A_{1}$, respectively. Furthermore it can be shown that the transfer probabilities are symmetric
$\gamma_{A_{1}^{a'};\mathcal{A}_{0}^{a}}=\gamma_{A_{0}^{a};\mathcal{A}_{1}^{a'}}$ (see SM)
and, considering normalization, there is only a single independent parameter,  which is taken to be
 $c=2\gamma_{A_{1}^{0};\mathcal{A}_{0}^{0}}-1$, among eight of them. The NDWU relation Eq.(\ref{g2}) in this case becomes (see SM)
 \begin{eqnarray}
 \langle A_0\rangle^{2}+\langle A_1\rangle^{2}+c^{2}-2c\langle A_0\rangle\langle A_1\rangle\leq 1. \label{qub}
\end{eqnarray}

\emph{Correlation boundaries from Local Properties --- } We now consider the simplest Bell  scenario in which there are two physical systems prepared in some joint state distributed to two space-like separated observers Alice and Bob. Each observer can choose to perform two alternative sharp measurements $\{A_{\nu}\}$ and $\{B_{\mu}\}$ ($\nu,\mu=0,1$)  on their local systems with two outcomes labeled by $a,b=0,1$, resulting a  joint probability distribution, or correlation,  $\{p(ab|\nu\mu)\}$. Furthermore we assume the non-signaling conditions, i.e., the local statistics of each observer does not depend on the measurement settings of  the other observer.

 When measurement settings are prefixed to be independent of any shared randomness the  correlations are referred to as pure correlation (PC).
When Alice and Bob  preshare some random variable,  the state they share as well as the measurements they choose may depend on this random variable. The correlations obtained in this way are referred to as mixed correlation (MC). For example, in quantum theory, PC are of form
\begin{eqnarray*}
p_{\rm pure}(ab|\nu\mu)=\operatorname{Tr}(\rho M_{A^{a}_{\nu}}\otimes M_{B^{b}_{\mu}})
\end{eqnarray*}
while  MC are of form
 \begin{eqnarray*}
p_{\rm mixed}(ab|\nu,\mu)=\sum_{\lambda}P(\lambda) \operatorname{Tr}(\rho_{\lambda} M^{\lambda}_{A^{a}_{\nu}}\otimes M^{\lambda}_{B^{b}_{\mu}})
\end{eqnarray*}
with preshared random variable $\lambda$ distributed according to some probability distribution $P(\lambda)$.
Quantum MC  can be generated  by   convex combinations of quantum PC \cite{ran}. We note that the set of PC in general might not be convex while the set of MC is by definition the convex hull of the set of all possible PC.  As a consequence the set of PC   determines the corresponding MC uniquely  but not vise versa so that PC encodes more information on the structure of underlying theory than MC. Also  the commonly tested correlation in laboratory is PC rather than MC.  Thus PC are more fundamental than mixed ones.  So far, all the correlation criteria such as  SDP method, NPA criterion, and the  criteria arising from principles such as Global Exclusive,  Macroscopic Locality and Local Orthogonality are intrinsic convex thus they are not cable of characterizing   PC \cite{ic,lo,cl}.

 Since two observers are space-like, the same joint probability $p(ab|\nu\mu)$ can be obtained by two different orders of measurements: Bob measures first or Alice measures first. Suppose that Bob or Alice measures first with measurement $B_\mu$ or $A_\nu$ yielding outcome $b$ or $a$ with probability $p({a|\nu})=\sum_bp(ab|\nu\mu)$ and $p({b|\mu})=\sum_ap(ab|\nu\mu)$, which are well defined due to non-signaling conditions, respectively. For the other observer, a conditional state $\omega_{b|\mu}$ or $\omega_{a|\nu}$ is  prepared accordingly, which determines the statistics of all the measurements concerned via the conditional probability
$$p_{\omega_{b|\mu}}({a|\nu})=\frac{p({ab|\nu\mu})}{p({b|\mu})},\quad
p_{\omega_{a|\nu}}({b|\mu})=\frac{p({ab|\nu\mu})}{p({a|\nu})}.$$

In the study of  general nonlocal correlations, it is a fundamental  question what correlation can arise from the subsystems exhibiting some local properties.     There has been some interesting results regarding to the understanding of  correlation from  perspectives of local properties:  any non-local theory ensures the local properties of intrinsic uncertainty, complementarity and  non-clone principle \cite{gs}; the violation of Clauser-Horne-Shimony-Holt (CHSH) is upper-bounded by a function of a balance strength characteristic for the underlying theory \cite{me}.  In following, we shall assume that the local system obeys the NDWU relation, i.e., in any given state (especially in those four effective conditional states) of each local system for any two sharp measurements it holds NDWU relation Eq.(\ref{qub}). This imposes strong constraints on the correlations:

{\bf Theorem 2} \emph{ In a non-signaling theory, if for each subsystem the NDWU relation Eq.(\ref{qub}) holds, then the  correlation $\{p(ab|\nu\mu)\}$  satisfy
\begin{eqnarray}
\max_{\mathcal S\in \mathfrak B}\left\{\langle A_0\rangle_{\mathcal S}\langle A_1\rangle_{\mathcal S}-\sqrt{1-\langle A_0\rangle_{\mathcal S}^2}\sqrt{1-\langle A_1\rangle_{\mathcal S}^2}\right\}
\nonumber\hskip 0.75cm\\
\leq\min_{\mathcal S\in \mathfrak B}\left\{\langle A_0\rangle_{\mathcal S}\langle A_1\rangle_{\mathcal S}+
\sqrt{1-\langle A_0\rangle_{\mathcal S}^2}\sqrt{1-\langle A_1\rangle_{\mathcal S}^2}\right\}, \label{g3a}\\
\max_{\mathcal S\in \mathfrak A}\left\{ \langle B_0\rangle_{\mathcal S}\langle B_1\rangle_{\mathcal S}-
\sqrt{1-\langle B_0\rangle_{\mathcal S}^2}\sqrt{1-\langle B_1\rangle_{\mathcal S}^2}\right\}\nonumber\hskip 0.75cm\\
\leq\min_{\mathcal S\in \mathfrak A}\left\{  \langle B_0\rangle_{\mathcal S}\langle B_1\rangle_{\mathcal S}+
\sqrt{1-\langle B_0\rangle_{\mathcal S}^2}\sqrt{1-\langle B_1\rangle_{\mathcal S}^2}\right\}, \label{g3b}
\end{eqnarray}
 where $\mathfrak{B}=\{\omega_{b|\mu}| b,\mu=0,1\}$, $\mathfrak{A} =\{\omega_{a|\nu}| a,\nu=0,1\}$, and}
\begin{eqnarray*}
\langle A_\nu\rangle_{\omega_{b|\mu}}=\frac{p({0b|\nu\mu})-p({1b|\nu\mu})}{p({0b|\nu\mu})+p(1b|\nu\mu)}=\frac{\langle A_\nu\rangle+(-1)^b\langle A_\nu B_\mu\rangle}{1+(-1)^b\langle B_\mu\rangle},\\
 \langle B_\mu\rangle_{\omega_{a|\nu}}=\frac{p({a0|\nu\mu})-p({a1|\nu\mu})}{p({a0|\nu\mu})+p({a1|\nu\mu})}=\frac{\langle B_\mu\rangle+(-1)^a\langle A_\nu B_\mu\rangle}{1+(-1)^a\langle A_\nu\rangle}.
 \end{eqnarray*}

The proof is presented in SM.  We note that the above result provides a quantum correlation criterion since the NDWU principle holds for QM.  We shall refer the constraints in Theorem 2 to as NDWU criterion. In the simplest Bell scenario QM violates the CHSH inequality and the maximum  violation is known as the  Tsirelson bound. As the first application we shall reproduce the Tsirelson bound:

{\bf Theorem 3 }\emph{In a non-signaling theory respecting the no disturbance without uncertainty principle it holds
\begin{equation}
\mbox{\rm CHSH}:=\sum_{a,b,\mu,\nu=0}^1(-1)^{a+b+\mu\nu}p(ab|\nu\mu)\leq2\sqrt2.
\end{equation}
}

\emph{Recovering quantum correlation boundary --- }
Any correlation violates Tsirelson's bound is not quantum mechanical and  there exist correlations that satisfy Tsirelson's bound but still not quantum mechanical.  This is not surprising since there are eight independent variables in the correlation and a single bound cannot fully characterize its behaviors.
 The NDWU criterion aims to characterize the boundaries of a general quantum correlation. We now use it to recover quantum boundary for a family of noisy PR-boxes.

 In the simplest Bell scenario, the full set of non-signaling  boxes  includes  eight extremal non-local boxes
 \begin{equation}\nonumber
 P^{\tau\sigma\lambda}_{NL}(ab|AB)=\left\{
 \begin{aligned}
 & \frac{1}{2} \quad \operatorname {if}  a\oplus b=AB\oplus \tau A \oplus \sigma B \oplus \lambda \\
  &0    \quad   \operatorname{otherwise},
 \end{aligned}
 \right.
 \end{equation}
 with $\tau, \sigma, \lambda \in \{0,1\}$, and  16 local deterministic boxes
 \begin{equation}\nonumber
 P^{\tau\sigma\lambda \varsigma}_{L}(ab|AB)=\left\{
 \begin{aligned}
 & 1 \quad \operatorname {if}  a= \tau A \oplus \sigma, \quad b=\lambda B \oplus \varsigma \\
  &0    \quad   \operatorname{otherwise}.
 \end{aligned}
 \right.
 \end{equation}
We consider the following 3-parameter family of noisy non-signaling box:
  \begin{eqnarray}\label{m1}
PR_{\alpha\beta\tau}=\alpha PR +\beta PR'+\tau L +\frac{I}{4}(1-\alpha- \beta-\tau).
\end{eqnarray}
 where $PR=P^{000}_{NL}$, $PR'=P^{010}_{NL}$, and $L=P^{0000}_{L}$.
From Tsirelson's bound we see that those correlations with parameters outside the region $\max\{4\alpha+2\tau, 4\beta+2\tau \}\leq 2\sqrt{2}$, which is  bounded by two blue planes in Fig.1a, cannot be quantum mechanical.

\begin{figure}
\includegraphics[scale=0.46]{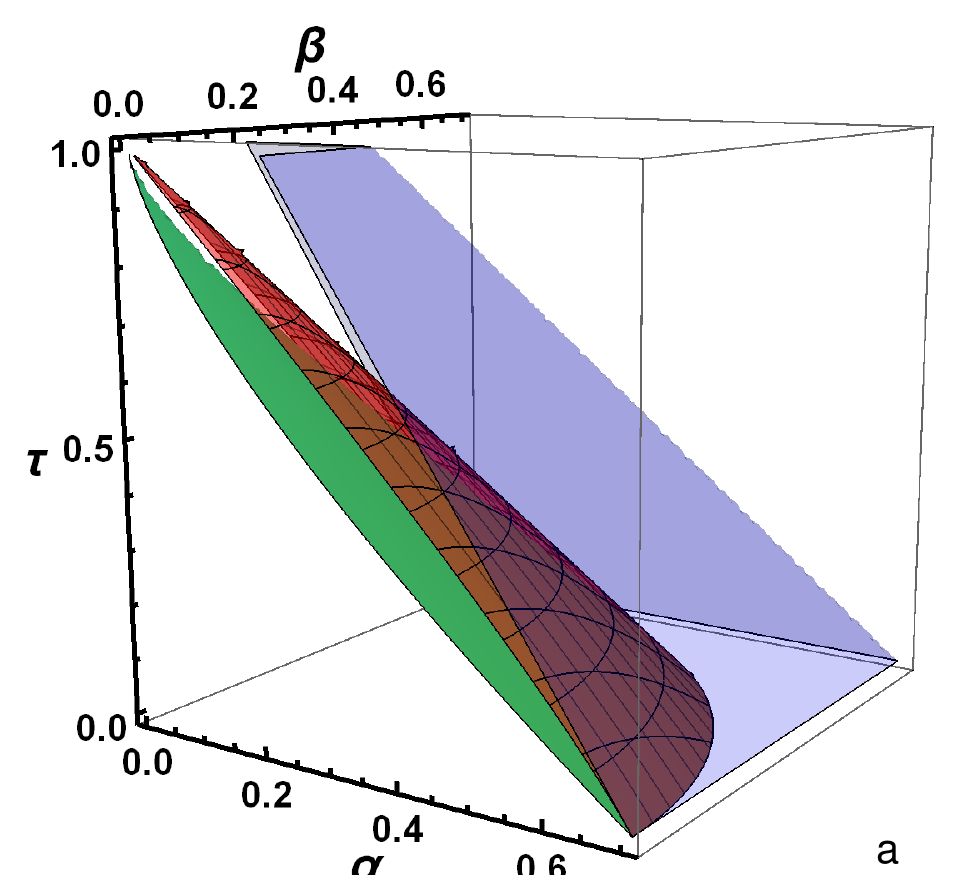}\ \includegraphics[scale=0.415]{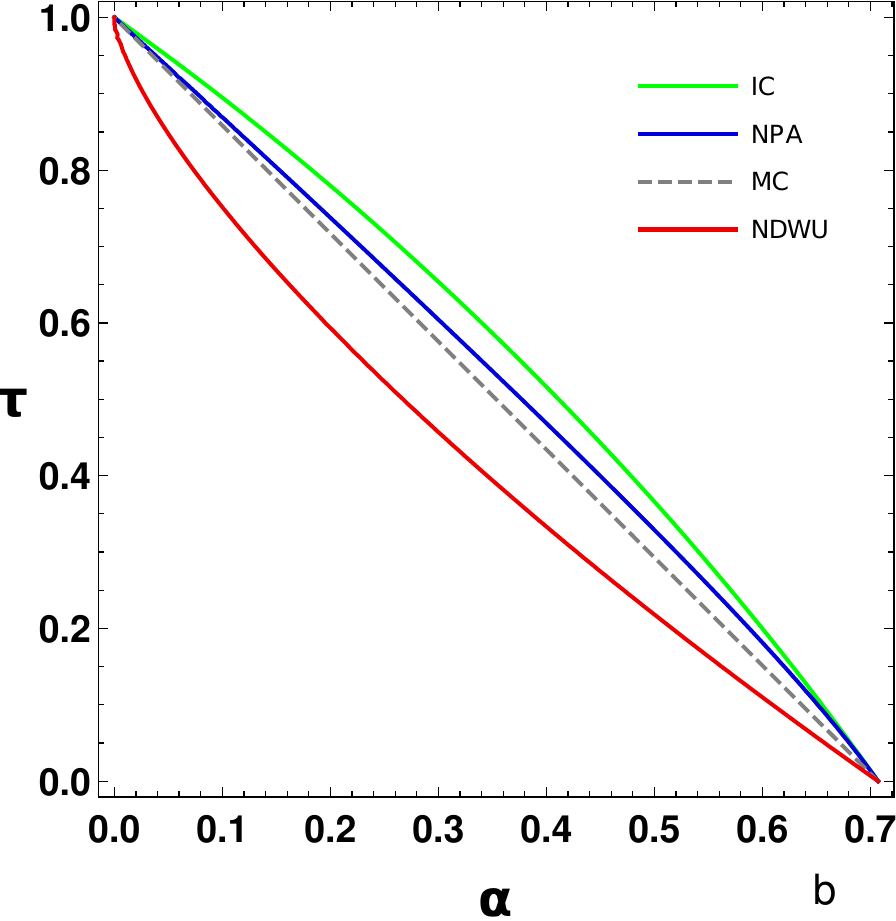}
\caption{(a) Boundaries of PC for 3-parametered family of non-signaling boxes implied by NDWU criterion Eq.(\ref{f}), by NPA criterion, and by Tsirelson's bound are represented by green surface, brown surface, and two blue planes, respectively. (b) The boundaries of PC for 2-parametered non-signaling box ($\beta=0$) implied by IC, NPA,  NDWU. The green  and blue solid curves indicate the boundaries from IC and NPA criterion, respectively, grey dashed line indicates the MC boundary from NDWU criterion, and the red curve represents the boundary of PC from the NDWU criterion.}
\end{figure}

We shall now apply our NDWU criterion provided by Theorem 2 to this family of correlations to derive  the  allowed regions of $\alpha$,  $\beta$, and $\tau$ for a quantum PC.
Two constraints   Eq.(\ref{g3a}) and Eq.(\ref{g3b}) end up with the following three-dimensional boundary (see SM)
\begin{widetext}
\begin{eqnarray}\label{f}
\sqrt{1-\left[\frac{\alpha+\beta}{1-\tau}\right]^{2}}\sqrt{1-\left[\frac{\alpha-\beta}{1-\tau}\right]^{2}}-
\frac{|\alpha^{2}-\beta^{2}|}{(1-\tau)^{2}}\hskip 8cm\nonumber\\
\geq \frac{(\alpha+\beta+2\tau)(|\alpha-\beta|+2\tau)}{(1+\tau)^{2}}
-
\sqrt{1-\left[\frac{\alpha+\beta+2\tau}{1+\tau}\right]^{2}}\sqrt{1-\left[\frac{|\alpha-\beta|+2\tau}{1+\tau}\right]^{2}},
     \end{eqnarray}
\end{widetext}
which is plotted in Fig.1a. We can see that NDWU criterion is strictly  tighter than the NPA criterion,  the most powerful   criterion we have until now.   In the case of  $\tau=0$, Eq.(\ref{f}) reduces to $\alpha^2+\beta^2 \leq\frac{1}{2}$, referred to as boundary 1 here, which coincides with the boundary given by NPA criterion and IC, being a sufficient quantum correlation boundary\cite{bou}. In the case of $\beta=0$, Eq.(\ref{f}) becomes
 $$\frac{(\alpha+2\tau)^{2}}{(1+\tau)^{2}}+\frac{\alpha^{2}}{(1-\tau)^{2}}\leq 1,$$ which is referred to as boundary 2, NPA criterion implies $$\left|3\arcsin\frac{\alpha+\tau-\tau^2}{(1-\beta^2)}-\arcsin\frac{\alpha-\tau-\tau^2}{1-\tau^2}\right|\leq \pi,$$ while  IC implies $(\alpha+\tau)^2+\tau^2 < 1$. These boundaries are compared in Fig.1b and it is clear that the most tight boundary is provided by our NDWU criterion.

Note that among all the criteria witnessing the quantum boundaries, only the boundary from NDWU criterion is non-convex, and it is the first PC criterion as far as we know.
   The  boundary 2 from NDWU criterion can generate  MC boundary as  $\beta+\sqrt{2}\alpha=1$ (by a convex mixing of  two extremal points $\{\beta=1, \alpha=0  \}$ and $\{\beta=0, \alpha=\frac{\sqrt{2} }{2} \}$ in boundary 2) as shown in Fig.1b.   The SDP method  is a convex criterion and shares  the two extremal points in this case, thus it would not give a tighter boundary than this MC boundary \cite{sdp}.  Regarding the powers of these methods in this case, the relation NDWU $\geq$ SDP $\geq$ NPA $\geq$  IC is shown in the figure.

\emph{Excluding Almost quantum correlation --- }The set of AQC  has been introduced to investigate to what extent the information-theoretic method is useful in bounding quantum correlation \cite{acq}.
  It has been found that AQC satisfies IC (with numeric evidence), and  all the current principles, thus these principles are fundamentally restricted  as they cannot exclude any AQC.

  An AQC that is not quantum mechanical and undetectable  even by the quantum NPA criterion \cite{acq,sdp} reads
  $$\left\{\frac{9}{20},\frac{2}{11},\frac{2}{11}, \frac{9}{20},\frac{22}{125}, \frac{\sqrt{2}}{9},\frac{37}{700}, \frac{22}{125}\right\}$$
which are eight probabilities of measuring observables $$(A_{0}, A_{1}, B_{0}, B_{1}, A_{0}B_{0}, A_{1}B_{0}, A_{0}B_{1}, A_{1}B_{1}),$$ respectively, with 1 as outcome. As there are  eight independent probabilities in the correlations in the simplest Bell scenario, we can recover all the correlations $\{p({ab|\nu\mu})\}$ so that the conditional expectation values $\{\langle A_\nu\rangle_{\omega_{b|\mu}}\}$ and $\{\langle B_\mu\rangle_{\omega_{a|\nu}}\}$ immediately. With a straightforward calculation, the right hand side of Eq.(\ref{g3a}) equals $-0.25$ while the left hand side equals $0.44$, i.e., the condition Eq.(\ref{g3a}) is violated, i.e., the AQC above is singled out from PC.
\begin{table}
\begin{center}
    \begin{tabular}{cccc}
\hline
Applications  & IC  & NPA & NDWU \\
\hline
Tsirelson's bound & Yes  & Yes & Yes\\
   Boundary 1 & Yes  & Yes & Yes\\
   Boundary 2 & No  & No  & Yes  \\
ACQ  & No  & No  & Yes      \\
 \hline
\end{tabular}
\end{center}
\caption{A comparison between IC, NPA, and NDWU criteria. }
\end{table}

We summarize the above comparison of NDWU criterion with  IC and NPA  criteria in Table I, showing that
our NDWU criterion outperforms both IC and NPA criteria. A fundamental question arises as whether NDWU criterion is a sufficient PC criterion.  Some evidences for  a positive answer  can be found in Ref \cite{quan}, where it is reported that if all the involved local systems and all the possible measurements performed on them allow a quantum mechanical description,   the correlations are also quantum. Differently,  the  correlation considered here deals with two dichotomic measurements only and four local conditional states on each side.  In this case NDWU principle is  sufficient to ensure  quantum description for these local measurements, from which it probably  follows that (no proof yet) NDWU criterion might also be a sufficient condition for quantum pure correlations.

Although our method has been demonstrated in details for the simplest Bell scenario, it should be emphasized that our method applies to most general scenario, e.g, multipartite and higher dimensional system with multi-settings.   The two cornerstones of our method, namely, the perspective  of understanding  correlation from  local properties of subsystems and the NDWU principle,  are applicable to general scenarios. Therefore our framework  could be taken as  a general method in characterizing  quantum correlations.

In conclusion, we have proposed {\it no disturbance without uncertainty} as a physical principle lying behind quantum theory.  As applications of this local principle to the non-signaling correlations, we have accounted for many quantum features such as the Tsirelson bound and  the quantum boundaries of a family of nonlocal boxes, and  ruled out  an almost quantum correlation from quantum correlation where other criteria all failed. In these applications, our NDWU criterion outperforms the celebrated quantum NPA criterion and  the previous information-theoretical results.

\section{Acknowledgement}

\onecolumngrid

\section{Supplemental Material}

\emph{Proof of  Theorem 1.--- } Suppose that a finite-level quantum system is prepared in  state  $\rho$ and two sharp measurements correspond to two orthonormal bases $\{|A^{a}_0\rangle\langle A^{a}_0|\}$ and $\{|A^{a'}_1\rangle\langle A^{a'}_1|\}$. By denoting $\langle A^{a}_{0}| A^{a'}_{1}\rangle=\Lambda_{a'a}$ we have  expansion
$|A^{a'}_{1}\rangle=\sum_{a}\Lambda_{a'a}|A^{a}_{0}\rangle$. As a result
$$p({a'|A_{1}})=\langle A^{a'}_{1}|\rho |A^{a'}_{1}\rangle=\sum_{s,t}\Lambda^{*}_{a's}\Lambda_{a't}\langle A^{s}_{0}|\rho| A^{t}_{0}\rangle=\sum_{s\neq t}\Lambda^{*}_{a's}\Lambda_{a't}\langle A^{s}_{0}|\rho| A^{t}_{0}\rangle+p({a'|A_{0}\to A_{1}})$$
where $p({a'|A_{0}\to A_{1}})=\sum_{a}p({a|A_{0}})\gamma_{A_{1}^{a'};\mathcal{A}_{0}^{a}}$ with $p(a|A_0)=\langle A_0^a|\rho |A_0^a\rangle$ and transfer probabilities are symmetric
$$\gamma_{A_{1}^{a'};\mathcal{A}_{0}^{a}}=\gamma_{A_{0}^{a};\mathcal{A}_{1}^{a'}}=|\langle A^{a'}_{1}|A^{a}_{0}\rangle|^{2}.$$
Finally we calculate
\begin{eqnarray*}
 D_{A_{0}\to A_{1}}&=& \sum _{a'}  \big|p({a'|A_{1}})-p({a'|A_{0}\to A_{1}})\big|\\
&=& \sum_{a'}  \big|\sum_{s\neq t}\Lambda^{*}_{a's}\Lambda_{a't}\langle A^{s}_{0}|\rho| A^{t}_{0}\rangle\big|\\
&\leq&\textstyle\sum_{a'}\sqrt{\sum_{s\neq t}|\Lambda^{*}_{a's}\Lambda_{a't}|^{2}}\sqrt{\sum_{s\neq t}|\langle A^{s}_{0}|\rho| A^{t}_{0}\rangle|^{2}}\\
&\leq&\textstyle \sum_{a'} \sqrt{\sum_{s\neq t}\gamma_{A_{0}^{t};\mathcal{A}_{1}^{a'}}\gamma_{A_{0}^{s};\mathcal{A}_{1}^{a'}}}\sqrt{\sum_{s\neq t}p({s|A_{0}})p({t|A_{0}})}\\
&=&\textstyle \sum _{a'} \sqrt{1-\sum_{s}\gamma_{A_{0}^{s};\mathcal{A}_{1}^{a'}}^{2}}\sqrt{1-\sum_{t}p({t|A_{0}})^{2}}\\
&=& \Delta_{A_{0}|A_{1}}\Delta_{A_{0}}.
\end{eqnarray*}
Here the first inequality is due to Cauchy inequality, the second inequality is due to the fact that the state is positive semidefinite so that in the basis $\{|A_0^a\rangle\}$ it holds
$|\langle A_0^s|\rho|A_0^t\rangle|^2\le \langle A_0^s|\rho|A_0^s\rangle \langle A_0^t|\rho|A_0^t\rangle$.

\emph{ The transfer probabilities are symmetric, i.e., $\gamma_{A^{a}_{0};\mathcal{A}^{a^\prime}_{1}}=\gamma_{A^{a^\prime}_{1};\mathcal{A}^{a}_{0}}$, for two-outcome sharp measurements if the NDWU relation holds for all pairs of sharp measurements --- } Considering normalization conditions, there are four independent transfer probabilities (which is independent of the initial state)
$$\gamma_{A^{0}_{1};\mathcal{A}^{0}_{0}}=\frac{1+c_1}2, \quad
\gamma_{A^{0}_{1};\mathcal{A}^{1}_{0}}=\frac{1+c_2}2,\quad
\gamma_{A^{0}_{0};\mathcal{A}^{0}_{1}}=\frac{1+c'_1}2,\quad
\gamma_{A^{0}_{0};\mathcal{A}^{1}_{1}}=\frac{1+c'_2}2.
$$
It is symmetric if and only if $c_1=-c_2=c_1'=-c_2'$.
By denoting the expectation values $\langle A_0\rangle_{\mathcal S}=p_{\mathcal S}({0|A_{0}})-p_{\mathcal S}({1|A_{0}})$ and
 $\langle A_1\rangle_{\mathcal S}=p_{\mathcal S}({0|A_{1}})-p_{\mathcal S}({1|A_{1}})$ for two sharp measurements in an arbitrary state $\mathcal S$, the NDWU relation
 Eq.(\ref{g2}) becomes
 $$
 \frac12\left(\sqrt{1-c_1'^2}+
\sqrt{1-c_2'^2}\right)\sqrt{1-\langle A_0\rangle_{\mathcal S}^2}\ge \left|\langle A_1\rangle_{\mathcal S}-\frac{c_1+c_2}2-\frac{c_1-c_2}2\langle A_0\rangle_{\mathcal S}\right|
 $$
 for the sequential measurement scenario $A_0\to A_1$ while
  $$
 \frac12\left(\sqrt{1-c_1^2}+
\sqrt{1-c_2^2}\right)\sqrt{1-\langle A_1\rangle_{\mathcal S}^2}\ge \left|\langle A_0\rangle_{\mathcal S}-\frac{c_1'+c_2'}2-\frac{c_1'-c_2'}2\langle A_1\rangle_{\mathcal S}\right|
 $$
  for the sequential measurement scenario $A_1\to A_0$. Let us take at first the state $\mathcal S=\mathcal A_1^0$, i.e., the state on which measurement $A_1$ gives a definite value $0$. In this case $\langle A_0\rangle_{\mathcal A_1^0}=c_1'$ and $\langle A_1\rangle_{\mathcal A_1^0}=1$ and noting that transfer probabilities are state independent, we obtain
  \begin{eqnarray*}
  1-\frac{c_1+c_2}2-\frac{c_1-c_2}2c_1'&\le&\frac12\left(\sqrt{1-c_1'^2}+
\sqrt{1-c_2'^2}\right)\sqrt{1-c_1'^2}\\
&\le&\sqrt{1-\left(\frac{|c_1'|+|c_2'|}2\right)^2}\sqrt{1-c_1'^2}\\
&\le&1-\frac{|c_1'|+|c_2'|}2|c_1'|\le 1-\frac{c_1'-c_2'}2c_1'
  \end{eqnarray*}
  where the first inequality is due to NDWU relation, the second inequality is due to the concavity of function $\sqrt{1-x^2}$, and the third inequality is due to inequality $\sqrt{1-x^2}\sqrt{1-y^2}\le 1-|xy|$.
 By taking the state to be $\mathcal S=\mathcal A_1^1$, i.e., the state on which measurement $A_1$ gives a definite value $1$, we have $\langle A_0\rangle_{\mathcal A_1^1}=c_2'$ and $\langle A_1\rangle_{\mathcal A_1^1}=-1$ so that
   \begin{eqnarray*}
  1+\frac{c_1+c_2}2+\frac{c_1-c_2}2c_2'&\le&\frac12\left(\sqrt{1-c_1'^2}+
\sqrt{1-c_2'^2}\right)\sqrt{1-c_2'^2}\\
&\le&\sqrt{1-\left(\frac{|c_1'|+|c_2'|}2\right)^2}\sqrt{1-c_2'^2}\\
&\le&1-\frac{|c_1'|+|c_2'|}2|c_2'|\le 1+\frac{c_1'-c_2'}2c_2'.
  \end{eqnarray*}
 As a result we obtain $$(c_1-c_2)(c_1'-c_2')\ge (|c_1'|+|c_2'|)^2\ge |c_1'-c_2'|^2\ge0.$$
 Similarly from the $A_1\to A_0$ scenario we obtain
 $$(c_1-c_2)(c_1'-c_2')\ge (|c_1|+|c_2|)^2\ge |c_1-c_2|^2.$$
 If $c_1'=c_2'$ or $c_1=c_2$ we have $c_1'=c_1'=c_1=c_2=0$ then we already have symmetric transfer probabilities. If $c_1\not= c_2$ and $c_1'\not= c_2'$ we obtain
 both $ c_1-c_2\ge c_1'-c_2'  $ and $ c_1'-c_2'\ge c_1-c_2  $, by noting they have the same sign, so that $ c_1-c_2= c_1'-c_2'  $. As a result we have both $c_1+c_2\ge0 $ and $c_1+c_2\le 0$ giving the desired symmetry property of transfer probabilities $c_1=-c_2=c_1'=-c_2'=c$. The NDWU relation becomes
 $$\sqrt{1-c^2}\sqrt{1-\langle A_0\rangle^2}\ge |\langle A_1\rangle-c\langle A_0\rangle|$$
which is exactly NDWU relation Eq.(\ref{qub}) for two-outcome sharp measurements.

\emph{ Proof of Theorem 2 --- } By explicitly showing the dependence on the state the NDWU relation Eq.(\ref{qub}) can be recast into the following equivalent form
$$\langle A_0\rangle_{\mathcal S}\langle A_1\rangle_{\mathcal S}-\sqrt{1-\langle A_0\rangle_{\mathcal S}^2}\sqrt{1-\langle A_1\rangle_{\mathcal S}^2}\le c\le \langle A_0\rangle_{\mathcal S}\langle A_1\rangle_{\mathcal S}+\sqrt{1-\langle A_0\rangle_{\mathcal S}^2}\sqrt{1-\langle A_1\rangle_{\mathcal S}^2}$$
Because $c$ is state-independent we shall have
$$\max_{\mathcal S}\left\{\langle A_0\rangle_{\mathcal S}\langle A_1\rangle_{\mathcal S}-\sqrt{1- \langle A_0\rangle_{\mathcal S}^2}\sqrt{1-\langle A_1\rangle_{\mathcal S}^2}\right\}\le
\min_{\mathcal S}\left\{\langle A_0\rangle_{\mathcal S}\langle A_1\rangle_{\mathcal S}+\sqrt{1-\langle A_0\rangle_{\mathcal S}^2}\sqrt{1-\langle A_1\rangle_{\mathcal S}^2}\right\}$$
to ensure the existence of $c$. In the simplest Bell scenario, two measurements with two outcomes from Bob's side prepares four conditional states $\mathfrak B=\{\omega_{b|\mu}\}$ for Alice while two measurements with two outcomes from Alice's side prepares four conditional states $\mathfrak A=\{\omega_{a|\nu}\}$ for Bob, which lead to the NDWU conditions in Theorem 2.

\emph{ Proof of Theorem 3 --- }
Note that NDWU relation Eq.(\ref{qub}) can be recast to an equivalent form
$$\frac{(\langle A_0\rangle+\langle A_1\rangle)^{2}}{2(1+c)}+\frac{(\langle A_0\rangle-\langle A_1\rangle)^{2}}{2(1-c)}\leq 1,$$
where $c=2\gamma_{A^{0}_{0};\mathcal{A}^{0}_{1}}-1$.
As a result
\begin{align} \nonumber
|\langle A_0\rangle_{\mathcal S}\pm\langle A_1\rangle_{\mathcal S}|\leq \sqrt{2(1\pm c)}
\end{align}
for an arbitrary state ${\mathcal S}$, especially for those four conditional states $\omega_{b|\mu}$ resulting from the measurement $B_\mu$ made by Bob with outcome $b$.
Now the Tsirelson bound can be derived:
 \begin{align} \nonumber
\operatorname{CHSH}&=\sum_{a,b,\nu,\mu=0}^1(-1)^{a+b+\mu\nu}p(ab|\nu\mu)\\ \nonumber
&=\sum_{a,b,\nu,\mu=0}^1(-1)^{a+b+\mu\nu}p(b|\mu)\frac{p(ab|\nu\mu)}{p(b|\mu)}\\
&=\sum_{b,\nu,\mu=0}^1(-1)^{b+\mu\nu}p(b|\mu)\langle A_\nu\rangle_{\omega_{b|\mu}}\\
&=\sum_{b,\mu=0}^1(-1)^{b}p(b|\mu)\Big(\langle A_0\rangle_{\omega_{b|\mu}}+(-1)^\mu\langle A_1\rangle_{\omega_{b|\mu}}\Big)\\
 &\leq \sum_{ b,\mu=0}^1p(b|\mu)\left|\langle A_0\rangle_{\omega_{b|\mu}}+(-1)^\mu\langle A_1\rangle_{\omega_{b|\mu}}\right| \\ \nonumber
&\leq \sum_{ b,\mu=0}^1p(b|\mu)\sqrt{2(1+(-1)^\mu c)} \\ \nonumber
&=\sum_{\mu=0}^1 \sqrt{2(1+(-1)^\mu c)}\leq 2\sqrt{2}.
\end{align}
where the footnote $\mu_{b}$ specify the conditional states yielding the expectations. The maximum in the fourth inequality is taken when $\gamma_{A^{0}_{0};\mathcal{A}^{0}_{1}}=\frac{1}{2}$, which is just the condition for the maximum violation in QM.

\emph{Proof of three dimensional boundary Eq.(\ref{f}) ---}
Let $${\rm PR}=\frac{1+(-1)^{a+b+\nu\mu}}4,\quad {\rm PR}^\prime=\frac{1+(-1)^{a+b+\nu \mu+\mu}}4,\quad {\rm L}=\frac{(1+(-1)^a)(1+(-1)^b)}4$$
the 3-parameter family of non-signaling box becomes
$$\alpha {\rm PR}+\beta {\rm PR}^\prime+\tau {\rm L}+\frac14(1-\alpha-\beta-\tau)=\frac{1+(-1)^aA_\nu+(-1)^bB_\mu+(-1)^{a+b}C_{\nu \mu}}4$$
with $\alpha,\beta,\tau\ge 0$ and $\alpha+\beta+\tau\le 1$ and
$A_\nu=B_\mu=\tau,$ $C_{\nu \mu}=(-1)^{\nu \mu}(\alpha+(-1)^{\mu}\beta)+\tau.$
Define
$$E_{\nu;\mu_{b}}=\frac{A_\nu+(-1)^bC_{\nu \mu}}{1+(-1)^b B_\mu},\quad  D^\pm_{\mu_{b}}=E_{0;\mu_{b}}E_{1;\mu_{b}}\pm\sqrt{1-(E_{0;\mu_{b}})^2}\sqrt{1-(E_{1;\mu_{b}})^2}$$
and
$$F_{\mu; \nu_{a}}=\frac{B_\mu+(-1)^aC_{\nu \mu}}{1+(-1)^aA_\nu},\quad d_{\nu_{a}}^\pm= F_{0; \nu_{a}}F_{1;\nu_{a}}\pm \sqrt{1-(F_{0;\nu_{a}})^2}\sqrt{1-(F_{1;\nu_{a}})^2}$$
We shall at first prove that the following boundary arises from NDWU constraint on Alice's side
\begin{equation}
\frac{(2\tau+\alpha+\beta)^2}{(1+\tau)^2}+\frac{(\alpha-\beta)^2}{(1-\tau)^2}\le 1
\end{equation}
and then prove that the boundary Eq.(\ref{f}) arises from NDWU constraints on Bob's side. Then we compare these two boundaries to arrive at Eq.(\ref{f}).
First, we consider the NDWU constraints from Alice's side.
\begin{itemize}
\item By definition we have
$$E_{0;\mu_{b}}=\frac{\tau+(-1)^b(\alpha+(-1)^\mu\beta+\tau)}{1+(-1)^b\tau},\quad E_{1;\mu_{b}}=\frac{\tau+(-1)^{b}((-1)^\mu\alpha+\beta+\tau)}{1+(-1)^b\tau}$$
or, by denoting $\alpha_\pm=\beta\pm\alpha$,
$$E_{\nu;0_{0}}=\frac{2\tau+\alpha_+}{1+\tau},\quad E_{\nu;0_{1}}=\frac{-\alpha_+}{1-\tau},\quad E_{\nu;1_{1}}=\frac{(-1)^\nu\alpha_-}{1-\tau},\quad E_{\nu;1_{0}}=\frac{2\tau-(-1)^\nu \alpha_-}{1+\tau}$$

\item Since $1=D_{0_{0}}^+=D_{0_{1}}^+=-D_{1_{1}}^+$, we have
$$D^+_m=\min\left\{ 1-2(E_{\nu;1_{1}})^2,D^+_{1_{0}}\right\},\quad D^-_M=\max\left\{2(E_{\nu;0_{0}})^2-1,2(E_{\nu;0_{1}})^2-1,D^-_{1_{0}}\right\}
$$
\item Lower bound $D_M=2(E_{\nu;0_{0}})^2-1$.
\begin{itemize}
\item From $(1+\tau)\alpha_+=(1-\tau)\alpha_++2\tau\alpha_+\le (1-\tau)(2\tau+\alpha_+)$ it follows
$2(E_{\nu;0_{0}})^2-1\ge 2(E_{\nu;0_{1}})^2-1$.
\item To prove
$ 2(E_{\nu;0_{0}})^2-1\ge D_{1_{0}}^-$
i.e.,
$$ 2(2\tau+\alpha_+)^2-(1+\tau)^2\ge 4\tau^2-\alpha_-^2-\sqrt{(1+\tau)^2-(2\tau+\alpha_-)^2}\sqrt{(1+\tau)^2-(2\tau-\alpha_-)^2}$$
or
$f(Z)\ge 0$ with $Z=(1+\tau)^2$, $a,b=2\tau\pm\alpha_-$ and
$$f(\nu)=\sqrt{\nu-a}\sqrt{\nu-b}-\nu+C, \quad C=2(2\tau+\alpha_+)^2-4\tau^2+\alpha_-^2$$
being a non-decreasing function of $\nu$ since
$$f^\prime(\nu)=\frac12\left(\frac{\sqrt{\nu-a}}{\sqrt{\nu-b}}+\frac{\sqrt{\nu-b}}{\sqrt{\nu-a}}\right)-1\ge 0.$$

We can assume
$ Z\ge C$, otherwise we have trivially $f(\nu)\ge0$. Thus we have $f(Z)\ge f(C)=\sqrt{C-a}\sqrt{C-b}\ge 0$.
We have only to check $C\ge a,b$, i.e.,
\begin{eqnarray*}
 2(2\tau+\alpha_+)^2-4\tau^2+\alpha_-^2- (2\tau\pm\alpha_-)^2=8\tau\alpha_++\alpha^2_+\mp 4\tau\alpha_-\ge 0
\end{eqnarray*}
which is obviously true since $\tau\ge 0$ and $\alpha_+\ge\alpha_-$.
\end{itemize}
\item Upper bound $D_m^+= D_{1_{0}}^+$, i.e.,
$$ (1+\tau)^2-2\lambda^2\alpha_-^2\le 4\tau^2-\alpha_-^2+\sqrt{(1+\tau)^2-(2\tau+\alpha_-)^2}\sqrt{(1+\tau)^2-(2\tau-\alpha_-)^2}$$
with $\lambda=\frac{1+\tau}{1-\tau}$. We can assume
$$\frac{1+\tau^2+6\tau}{(1-\tau)^2}\alpha_-^2=(2\lambda^2-1)\alpha_-^2\le (1+\tau)^2-4\tau^2=(1-\tau)(1+3\tau)$$
i.e., we have $(1-\tau)^3\ge\alpha_-^2$.
Compute
$$\Big((1+\tau)^2-(2\tau+\alpha_-)^2\Big)\Big((1+\tau)^2-(2\tau-\alpha_-)^2\Big)-\Big( (1+\tau)^2-2\lambda^2\alpha_-^2+\alpha_-^2-4\tau^2\Big)^2$$
$$=(1+\tau)^2\Big(2(2\lambda^2-1)\alpha_-^2+8\tau^2-8\tau^2-2\alpha_-^2\Big)-4\lambda^2\alpha_-^2\Big(\lambda^2\alpha_-^2-\alpha_-^2+4\tau^2\Big)$$
$$=4(1+\tau)^2(\lambda^2-1)\alpha_-^2-4\lambda^2\alpha_-^2\Big(\lambda^2\alpha_-^2-\alpha_-^2+4\tau^2\Big)$$
$$=4\alpha_-^2\Big((\lambda^2-1)((1+\tau)^2-\lambda^2\alpha_-^2)-4\lambda^2\tau^2\Big)=\frac{16\tau\alpha_-^2\lambda^2}{(1-\tau)^2}((1-\tau)^3-\alpha_-^2)\ge 0$$
\end{itemize}

And then we consider the NDWU constraints from Bob's side. We assume $\alpha_-\ge 0$ i.e., $\beta\ge \alpha$ in what follows. By definition we have
$$F_{\mu;\nu_{a}}=\frac{\tau+(-1)^a(\tau+(-1)^{\nu\mu}(\alpha+(-1)^\mu\beta))}{1+(-1)^a\tau}$$
leading to
$$F_{0;\nu_{0}}=\frac{2\tau+\alpha_+}{1+\tau}:=p,\quad F_{0;\nu_{1}}=\frac{-\alpha_+}{1-\tau}:=-u,$$
and
$$ F_{1;0_{0}}=\frac{2\tau-\alpha_-}{1+\tau}:=q,\quad F_{1;1_{0}}=\frac{2\tau+\alpha_-}{1+\tau}:=q^\prime, \quad F_{1;0_{1}}=\frac{\alpha_-}{1-\tau}=v,\quad F_{1;1_{1}}=\frac{-\alpha_-}{1-\tau}=-v.$$
First we note that function $f(x,y)=xy-\sqrt{1-x^2}\sqrt{1-y^2}$ is non-decreasing for $x,y\ge0$ since
$$\partial_x f(x,y)=y+x\frac{\sqrt{1-y^2}}{\sqrt{1-x^2}}\ge0,\quad \partial_y f(x,y)=x+y\frac{\sqrt{1-x^2}}{\sqrt{1-y^2}}\ge0$$
and the function $g(x,y)=xy+\sqrt{1-x^2}\sqrt{1-y^2}$  is non-decreasing for $x\ge y$ since
$$\partial_y g(x,y)=\sqrt{1-x^2}\left(\frac x{\sqrt{1-x^2}}-\frac y{\sqrt{1-y^2}}\right)\ge 0\quad (x\ge y)$$

\begin{itemize}
\item Lower bound: $\max\{d^-_{\nu_{a}}\}=d^-_{1_{0}}$ because
\begin{itemize}
\item $d^-_{1_{1}}=f(u,v)\ge f(-u,v)= d^-_{0_{1}}$ since $u,v\ge 0$ as $\alpha_-\ge 0$.
\item $d^-_{1_{0}}=f(p,q^\prime)\ge d^-_{0_{0}}=f(p,q)$.
\item $d^-_{1_{0}}=f(p,q^\prime)\ge f(p,v) \ge f(u,v)=d^-_{1_{1}}$
because $p\ge u$ and $q^\prime\ge v$ as
$$(1-\tau)(2\tau+\alpha_\pm)-(1+\tau)\alpha_\pm=2\tau(1-\tau-\alpha_\pm)\ge 0$$
\end{itemize}
\item Upper bound: $\min\{d^+_{\nu_{a}}\}=d^+_{0_{1}}$
\begin{itemize}
\item $d^+_{1_{1}}=g(u,v)\ge g(-u,v)= d^+_{0_{1}}$
\item $d^+_{1_{0}}=g(p,q^\prime)\ge g(p,q)= d^+_{0_{0}}$ since $p\ge q,q^\prime$
\item Need to show $d^+_{0_{0}}\ge d^+_{0_{1}}$ i.e.,
$ g(p,q)=pq+\bar p\bar q\ge -uv+\bar u\bar v:=c$.
As a result $\Delta:= x^2+y^2-2 x y c=2(1-c)( x+ y)-(1-c)^2$ where
$$t=\frac{1-\tau}{1+\tau},\quad  x=1-u,\quad  y=1+v, \quad p=1-t  x,\quad q=1-t  y$$
Then either $c\le pq$ which gives trivially the result or $c\ge pq$ in which case $g(p,q)\ge c$ becomes
$$0\ge p^2+q^2+c^2-1-2pqc=t^2\Delta-2t(1-c)( x+ y)+(1-c)^2$$
or equivalently
$$t_c=\frac{1-c}{2 x+2 y-1+c}\le t\le 1$$
which is ensured by $c\ge pq=1-t( x+ y)+t^2 x y$, i.e., $t_-\le t\le t_+$ where
$$t_\pm=\frac{ x+ y\pm\sqrt{( x- y)^2+4 x y c}}{2 x  y}$$
as long as we can show
$t_c\le t_-$. Noting that
$$ x+ y-1+c= x+ y-1-(1- x)( y-1)+\bar u\bar v= x y+\bar u\bar v\ge 0$$ we have
$$\frac 1{t_c}=\frac{2 x+2 y}{1-c}-1\ge\frac{ x+ y}{1-c}\ge\frac{ x+ y+\sqrt{( x+ y)^2-4 x y(1-c)}}{2(1-c)}=\frac1{t_-}.$$

\end{itemize}

In sum, we have $ d^+_m=d^+_{0_{1}}$ and $d^-_M=d^-_{1_{0}}$ in the case of $\alpha_-\ge0$. Similarly we have $ d^+_m=d^+_{1_{0}}$ and $d^-_M=d^-_{0_{1}}$ in the case of $\alpha_-\le 0$ which gives rise to bound Eq.(\ref{f}).

\begin{figure}
\includegraphics[scale=0.4]{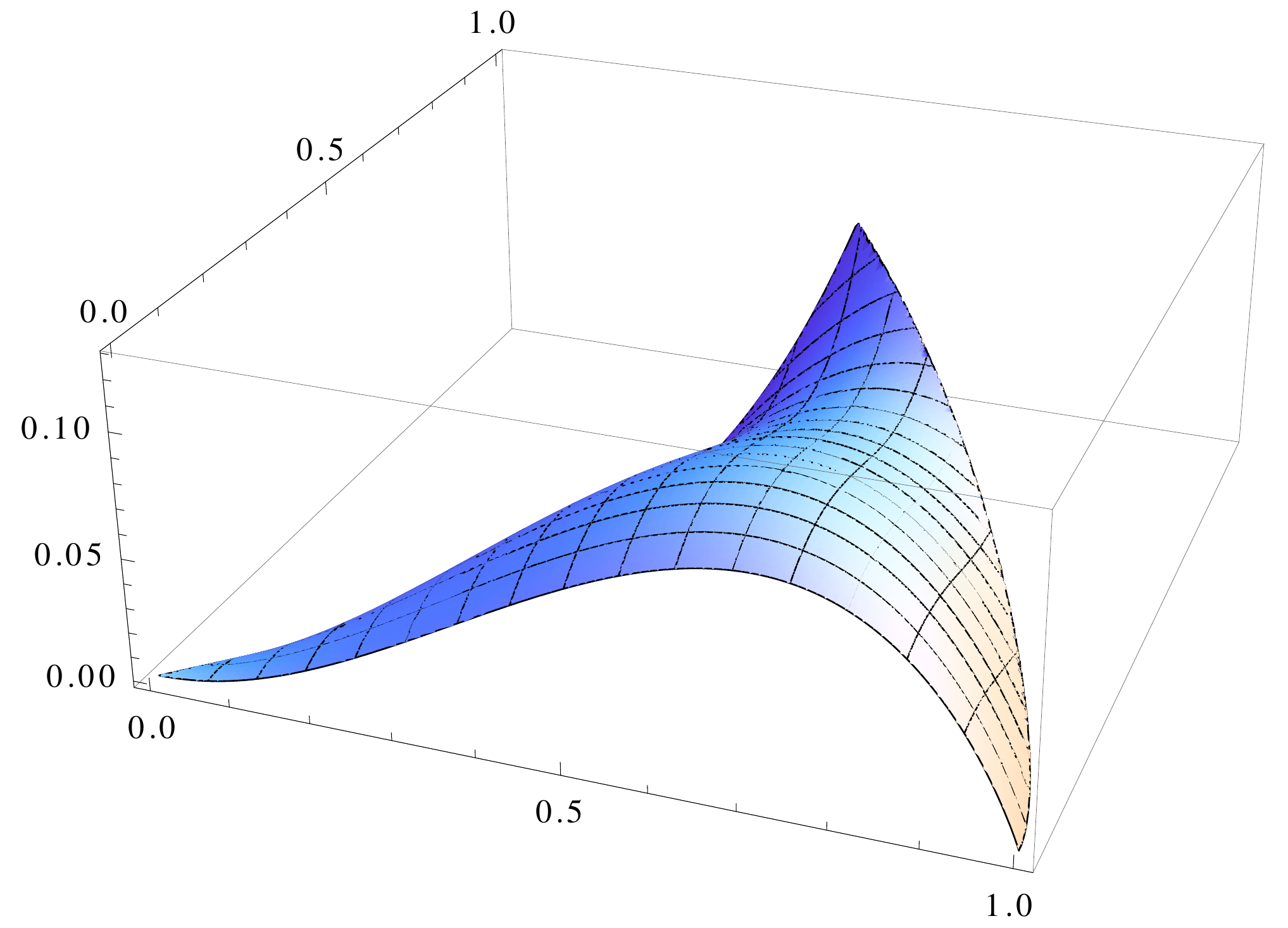}
\caption{A plot of $(1-c)(x+1-\bar v)-2(x+z)(1-\bar v)$ as a function of $u=1-x$ and $v$ in the region of $u\ge v$ and $u+v\le 1+c$.}
\end{figure}

\end{itemize}

Last,  we shall compare two boundaries arising from Alice and Bob. Assume still $\alpha_-\ge0$ and
we shall prove that Alice's boundary $d_{0_{1}}^+\ge d_{1_{0}}^-$, i.e., $\bar p\bar q^\prime+\bar u\bar v-p q^\prime- uv\ge 0$ infers Bob's boundary $p^2+v^2\le1$ or $1\ge t\ge (1-\bar v)/ x$. We have $q^\prime=1-tz$ with $z=2- y$ ($z\ge  x$) so that $v= y-1=1-z$. In terms of $z$ we have
 again  either  \begin{itemize}
 \item $\bar u\bar v-uv=c\ge pq^\prime$, i.e., $c\ge (1-tx)(1-tz)=1-t(x+z)+t^2xz$. If $(x+z)^2<4(1-c)xz$ then we have $c<pq^\prime$. Thus we should have $(x+z)^2\ge 4(1-c)xz$ giving rise to
 $$T_+=\frac{x+z+\sqrt{(x+z)^2-4(1-c)xz}}{2xz}\ge t\ge \frac{x+z-\sqrt{(x+z)^2-4(1-c)xz}}{2xz}:=T_-$$
  or
 \item $c\le pq^\prime$ with
$p^2+{q^\prime}^2+c^2-2pq^\prime c\le 1$.
 i.e., $t\ge T_+$ or $t\le T_-$, and
$$0\ge(1-c)^2-2t( x+z)(1-c)+t^2( x^2+z^2-2 x zc)$$
i.e.,
$$t_p:=\frac{1-c}{ x+z-\sqrt{2(1+c) x z}}\ge t \ge \frac{1-c}{ x+z+\sqrt{2(1+c) x z}}:=t_m$$
\item From $c=\bar u\bar v-u v$ it follows that
$$|1-c-x-z|=\sqrt{2(1+c)xz}$$
leading to $t_m=1$ if $1-c\ge x+z$ (trivially infering Bob's boundary) and $t_p=1$ if $1-c\le x+z$.
\end{itemize}

Since
$$t_m=\frac{1-c}{ x+z+\sqrt{2(1+c) x z}}\le\frac{2(1-c)}{2(x+z)}\le \frac{2(1-c)}{ x+z+\sqrt{(x+z)^2-4(1-c) x z}}= T_-$$ we obtain finally Alice's boundary as $$1\ge t\ge  t_m\cap 1-c\le x+z.$$
We have only to show that,  given $1-c\le x+z$,
$$t_m=\frac{1-c}{ 2(x+z)-1+c}\ge \frac{1-\bar v} x$$
or
$$(1-c)(x+1-\bar v)-2(x+z)(1-\bar v)\ge0$$
which is plotted in Fig.2 as a function of $u$ and $v$ which is clearly non-negative.


\begin{thebibliography}{99}
%-------------------------------------------------------------------------

\bibitem{bel}   N. Brunner, D. Cavalcanti, S. Pironio, V. Scarani,   and S. Wehner, Bell nonlocality   Rev. Mod. Phys. \textbf{86}, 419 (2014).


\bibitem{npa}    M.Navascues, S. Pironio, and A. Ac\'{\i}n, Bounding the set of quantum correlations, 	Phys. Rev. Lett. \textbf{98}/ , 010401 (2007).
\bibitem{tir} B. S. Cirel'son, Lett. Math. Phys. \textbf{4}, 93 (1980).
\bibitem{sp} S. Popescu,  Nonlocality beyond quantum mechanics,   Nat. Phys \textbf{10}, 264¨C270 (2014).
\bibitem{ic}  M. Pawlowski, T. Paterek, D. Kaszlikowski, V. Scarani, A. Winter, and M. Zukowski. Information Causality as a Physical Principle.  \emph{Nature} \textbf{461}, 1101 (2009).
\bibitem{wvd} W. van Dam. Implausible consequences of superstrong nonlocality. \emph{Nat. Comput} 12:9-12 (2013).
\bibitem{cl} A. Cabello.  Simple Explanation of the Quantum Violation of a Fundamental Inequality. \emph{Phys. Rev. Lett.} \textbf{110}, 060402 (2013).
\bibitem{by} B. Yan. Quantum Correlations are Tightly Bound by the Exclusivity Principle. \emph{Phys. Rev. Lett.} \textbf{110}, 260406 (2013).
 \bibitem{ca}  A. Cabello. Exclusivity principle and the quantum bound of the Bell inequality. \emph{Phys. Rev. A} \textbf{90}, 062125 (2014).
\bibitem{lv}  L. Vandenberghe and S. Boyd, SIAM Rev. \textbf{38}, 49 (1996).
\bibitem{sdp1}  L. Masanes,  arxiv:quant-ph/0309137.
    \bibitem{sdp}   M. Navascues, S. Pironio, and A. Ac\'{\i}n,   New J. Phys. \textbf{10}, 073013 (2008).
    \bibitem{dim} M. Navascu\'{e}s, and T. VertesiPhys, Phys. Rev. Lett. \textbf{115}, 020501 (2015).
\bibitem{hh}   S. Ishizaka, Phys. Rev. A \textbf{97}, 050102.




     \bibitem{tem} C. Budroni, T. Moroder, M. Kleinmann, and O. G\"{u}hne, Phys. Rev. Lett. \textbf{111}, 020403 (2013).
  \bibitem{acq} M. Navascu\'{e}s, Y. Guryanova, M. J. Hoban,  and A. Ac\'{\i}n, Nat. Commun \textbf{6}, 6288 (2015).
  \bibitem{pr}   S. Popescu  and D, Rohirlich. Quantum nonlocality as an axiom. \emph{Found. Phys.} \textbf{24} 379 (1994).
   \bibitem{tn1} S. Popescu,   Nonlocality beyond quantum mechanics.  \emph{Nature Phys}, \textbf{10}, 264 (2014).


 \bibitem{ml} M. Navascu\'{e}s and H. Wunderlich, Proc. Royal Soc. A \textbf{466}:881-890 (2009).
  \bibitem{lo}  T. Fritz, A. B. Sainz, R. Augusiak, J. B. Brask, R. Chaves, A. Leverrier and A. Ac\'{\i}n, arXiv:1210.3018.
   \bibitem{me}Liang-Liang Sun, Sixia Yu, Zeng-Bing Chen, arXiv:quant-ph/ 1808.06416.
     \bibitem{ran} J. M. Donohue and E. Wolfe, Phys. Rev. A \textbf{92}, 062120.
  \bibitem{gpt}  P. Janotta, and H. Hinrichsen.  Generalized probability theories: what determines the structure of quantum theory? \emph{J. Phys. A: Math. Theor.} \textbf{47},  323001 (2014).

   \bibitem{gs} Li. Masanes, A. Acin, and N. Gisin. General properties of nonsignaling theories. \emph{Phys. Rev. A} \textbf{73}, 012112 (2006).




     \bibitem{15'}  T. V\'{e}rtesi and N. Brunner, Nat. Commun. \textbf{5}, 5297 (2014).
 \bibitem{16'}E. Wolfe and S. F. Yelin, Phys. Rev. A \textbf, 012123 (2012).
 \bibitem{17'} K. T. Goh, J. Kaniewski, E. Wolfe, T. V\'{e}rtesi, X. Wu, Y. Cai,
Y.-C. Liang, and V. Scarani, Phys. Rev. A \textbf{97}, 022104 (2018).





    \bibitem{bou}          J. Allcock, N. Brunner, M. Paw{\l}owski, and V. Scarani, Recovering part of the boundary
 between quantum and nonquantum correlations from information causality. Phys. Rev. A \textbf{80}, 040103 (2009).
 \bibitem{quan} H. Barnum, S. Beigi, S. Boixo, M. B. Elliott, and S. Wehner, Phys. Rev. Lett. 104, 140401.











%---------------- -----------------------------------------------------------
\end{thebibliography}
\end{document}